\documentclass[12pt,preprint]{aastex}

\newcommand{\HI}{H~{\sc i}} 

\newcommand{\kms}{${\rm km~s^{-1}}$}

\shortauthors{MCCLURE-GRIFFITHS ET AL} 
\shorttitle{Strands of Cold Hydrogen}

\begin{document} 

\title{Magnetically Dominated Strands of Cold Hydrogen in the Riegel-Crutcher Cloud}

\author{N.\ M.\ McClure-Griffiths,\altaffilmark{1} J.\ M.\
  Dickey,\altaffilmark{2} B.\ M.\ Gaensler,\altaffilmark{3,4} A.\ J.\
  Green,\altaffilmark{5} and Marijke Haverkorn\altaffilmark{6,7} }

\altaffiltext{1}{Australia Telescope National Facility, CSIRO, PO Box 76, Epping
 NSW 1710, Australia; naomi.mcclure-griffiths@csiro.au} 
\altaffiltext{2}{School of Mathematics and Physics,
 University of Tasmania, Private Bag 21, Hobart TAS 7001, Australia; john.dickey@utas.edu.au}
\altaffiltext{3}{Alfred P.\ Sloan Fellow; Harvard-Smithsonian Center for Astrophysics, 60 Garden Street, Cambridge, MA 02138}
\altaffiltext{5}{Current Address: School of Physics, The University of
  Sydney A29, NSW 2006, Australia; bgaensler@physics.usyd.edu.au}
\altaffiltext{5}{School of Physics, The University of Sydney A29, NSW 2006, Australia; agreen@physics.usyd.edu.au}
\altaffiltext{6}{Astronomy Department, University of California, 601 Campbell Hall, Berkeley, CA 84720; marijke@astro.berkeley.edu}
\altaffiltext{7}{Jansky Fellow, National Radio Astronomy Observatory}
\authoraddr{Address correspondence regarding this manuscript to: 
                N. M. McClure-Griffiths
                ATNF, CSIRO
                PO Box 76
		Epping NSW 1710
		Australia }
\begin{abstract}
  We present new high resolution ($100\arcsec$) neutral hydrogen (\HI)
  self-absorption images of the Riegel-Crutcher cloud obtained with
  the Australia Telescope Compact Array and the Parkes Radio
  Telescope. The Riegel-Crutcher cloud lies in the direction of the
  Galactic center at a distance of $125\pm25$ pc.  Our observations
  resolve the very large, nearby sheet of cold hydrogen into a
  spectacular network of dozens of hair-like filaments.  Individual
  filaments are remarkably elongated, being up to 17 pc long with
  widths of less than $\sim 0.1$ pc.  The strands are reasonably cold,
  with spin temperatures of $\sim 40 $ K and in many places appearing to
  have optical depths larger than one.  Comparing the \HI\ images with
  observations of stellar polarization we show that the filaments are
  very well aligned with the ambient magnetic field.  We argue that
  the structure of the cloud has been determined by its magnetic
  field. In order for the cloud to be magnetically dominated the
  magnetic field strength must be $>30~{\rm \mu G}$.

\end{abstract}
\section{Introduction}
\label{sec:intro}
Recent high resolution surveys of neutral hydrogen (\HI) in the
Galactic Plane show that the structure of the warm atomic medium
varies significantly on arcminute scales.  The detailed structure of
cold \HI, however, is much more elusive.  Most of our knowledge of
cold \HI\ comes from observations of \HI\ absorption towards continuum sources
\citep[e.g.][]{heiles03b,heiles05}.  Although these measurements are
effective for determining gas temperature, they have limited
applicability to structural studies because they only trace gas along
single lines of sight towards an irregularly distributed array of
background sources. Another probe of cold \HI\ is \HI\ self-absorption
(HISA) towards bright background \HI\ emission.  This is not, strictly
speaking, self-absorption, but absorption of background \HI\ emission
by cold foreground \HI.  Surveys like the Canadian Galactic Plane
Survey \citep[CGPS;][]{taylor03} and the Southern Galactic Plane
Survey \citep[SGPS;][]{mcgriff05} have revived studies of HISA,
showing that it is a good probe of the structure of cold \HI\ on small
scales \citep[e.g.][]{gibson00,kavars05}.  Although these studies are
able to image HISA over large areas, and have better estimates of the
background emission than lower resolution surveys, it remains
difficult to unravel the structure of the cold foreground \HI\ from
variations in the background emission.

One of the most famous examples of a large scale HISA feature is the
Riegel-Crutcher \citep[hereafter R-C;][]{riegel72} cloud in the
direction of the Galactic center.  The R-C cloud was first discovered
by \citet{heeschen55} in his survey of \HI\ at the Galactic center and
he suggested that the cloud was very extended.  Subsequent
observations by \citet{riegel69} showed that the cold cloud extends
over $\sim 40$ degrees of Galactic longitude and $\sim 10$ degrees of
latitude.  It has even been suggested that the cloud is part of a much
larger structure, Lindblad's Feature A, spanning the full 360 deg of
Galactic longitude \citep{lindblad73}.  Estimates of the temperature
of the cloud show that despite its striking appearance it is not
exceptionally cold, with most estimates of spin temperature ranging
between 35 - 45 K \citep{montgomery95,riegel72,riegel69,heeschen55}.
By measuring \ion{Ca}{2} lines in the spectra of OB stars
\citet{crutcher74} determined that the cloud is between 150 and 180 pc
distant.  Further \ion{Na}{1} observations by \citet{crutcher84}
helped constrain the distance to $125\pm25$ pc, with an estimated
thickness of 1 to 5 pc.

The R-C cloud provides us with one of the best opportunities for
imaging the cold neutral medium (CNM) over a large area.  The \HI\
background towards the center of the Galaxy is amongst the brightest
anywhere in the Galaxy; this direction provides an ideal bright,
high column density background against which to image the CNM.  Here
we present new high resolution ($\sim 100\arcsec$) \HI\ images of the
R-C cloud, which allow us to study the structure of the CNM on scales
of $0.06$ pc to 22 pc.  In this paper we explore the relationship of
the R-C cloud to the ambient magnetic field and discuss its
implications for the structure of the CNM throughout the Milky Way.

This paper is structured as follows: in \S \ref{sec:obs} we discuss
the data taking and reduction strategy used.  In \S \ref{sec:results}
we present images of the R-C cloud and in \S \ref{subsec:tempstruc} we
derive optical depth and column density maps.  We present evidence of
an associated magnetic field in \S \ref{sec:bfield}. In \S
\ref{subsec:bstrength} we discuss the strength of the magnetic field and
its effect on the structure of the cloud.  In \S \ref{subsec:cnm} and
\ref{subsec:molecules} we comment on the structure of the R-C cloud,
its similarity to molecular clouds and the implications for models for
the structure of the CNM.

\section{Observations and Data Analysis}
\label{sec:obs} 
The data presented here were obtained as part of an extension to the
Southern Galactic Plane Survey \citep[SGPS;][]{mcgriff05}.  Between
late 2002 and early 2004 the SGPS was extended to cover $100~{\rm
  deg}^2$ around the Galactic Center.  The SGPS Galactic Center (GC)
survey, like the main SGPS, is a survey of the 21-cm continuum and
\HI\ spectral line emission in large regions of the Galactic Plane.
The GC survey covers $-5\arcdeg \leq l \leq +5\arcdeg$ and $-5\arcdeg
\leq b \leq +5\arcdeg$.  The main goal of the SGPS GC survey is to
study the structure and dynamics of \HI\ in the inner $\sim 1$ kpc of
the Galaxy, a topic which we will defer to a future paper.

The GC survey combines data from the Australia Telescope Compact Array
(ATCA) and the Parkes Radio Telescope for full sensitivity to angular
scales from 10 degrees down to the $100\arcsec$ resolution of the
data.  The Parkes data were observed and imaged as part of the main
SGPS survey and are fully described in \cite{mcgriff05}. The ATCA data
were taken in a slightly different manner than the data for the rest
of the SGPS.  For completeness we describe the full ATCA GC survey
parameters here.

The ATCA is an interferometer of six 22 m dishes situated near
Narrabri, New South Wales, Australia.  ATCA data were obtained during
eight observing sessions between 2002 December and 2004 June.  An
additional day of make-up observations was scheduled in 2004 November
to fill in gaps in the {\em u-v} coverage.  The ATCA has linear feeds
that receive two orthogonal linear polarizations, $X$ and $Y$, and can
observe at two intermediate frequencies (IFs) simultaneously.  As in
the SGPS, all data were recorded using a correlator configuration that
records the autocorrelations, $XX$ and $YY$, in 1024 channels across a
4 MHz bandwidth, as well as the full polarization products, $XX$,
$YY$, $XY$, and $YX$, in 32 channels across a 128 MHz bandwidth.  The
first IF was centered at 1420 MHz to observe the \HI\ line and the
second was centered at 1384 MHz.  Here we describe only the
narrow-band \HI\ data and reserve discussion of the continuum data for
a future paper.

Observations were made using six east-west array configurations with
maximum baselines between 352 m and 750 m.  Most baselines, in
multiples of 15 m, between 31 m and 750 m were sampled by the
combination of the configurations EW352, EW367, 750A, 750B, 750C, and
750D.  The observing arrays and dates are given in
Table \ref{tab:obs}.  The 100 ${\rm deg^2}$ field was covered by a
total of 967 individual pointings; 948 pointings were newly observed
and the remaining 19 pointings were taken from the SGPS Phase I.  All
pointings were distributed on a common hexagonal pattern with a separation of
$19\arcmin$ between adjacent pointings.  The observations were made in
snap-shot mode with 60 s integrations per pointing.  Each pointing was
observed approximately 30 times for a total integration time of $\sim
30$ min per pointing.

The primary flux calibrator, PKS B1934-638, was observed once per day
for bandpass and absolute flux calibration, assuming a flux at 1420
MHz of 14.86 Jy \citep{reynolds94}.  A secondary calibrator, PKS
B1827-360, was observed approximately every 45 minutes for complex
gain and delay calibration.  Data editing and calibration were
completed within the MIRIAD data reduction package using standard
techniques \citep{sault04}.

The 967 pointings were linearly combined and imaged using a standard
grid-and-FFT scheme.  The mosaicing process uses the joint approach,
where dirty images are linearly combined and jointly deconvolved
\citep{sault96}.  The joint imaging and deconvolution process improves
the sensitivity to large angular scale structures, increasing the
maximum scale sampled from $\sim 23\arcmin$ for a single pointing
observation to $\sim 30\arcmin$ for the mosaiced image
\citep{mcgriff05}. The deconvolution uses a maximum entropy method,
which is very effective at deconvolving large-scale emission.  The
maximum entropy algorithm in MIRIAD does not deconvolve the dirty beam
from areas of ``negative emission'', such as continuum sources
observed in absorption.  The Galactic center region contains many very
strongly absorbed continuum sources that would not be deconvolved in
standard processing.  To reduce the confusing sidelobe levels observed
around these sources we modeled the sources and subtracted them from
the {\em u-v} data prior to imaging.  Although we were not able to
completely remove the sources, the resultant sidelobes were
significantly reduced and do not adversely affect the analysis of the
\HI\ cube presented here.

Finally, the deconvolved ATCA image cube was combined with the Parkes
data cube to recover information on angular scales larger than $\sim
36\arcmin$.  This combination was performed using the MIRIAD task
IMMERGE, which Fourier transforms the deconvolved ATCA and Parkes
images, reweights them such that the Parkes image is given more weight
for large angular scales and the ATCA image more weight for small
angular scales.  The weighted, Fourier transformed images are linearly
combined and then inverse Fourier transformed \citep{stanimirovic02}.
The final datacube has a resolution of $100\arcsec$ and is sensitive
to angular scales up to the $10\arcdeg$ image size.  The rms
brightness temperature sensitivity in the combined cube is $\sim 2$ K.
The rms increases to $\sim 3$ K in the $\sim 1~{\rm deg^2}$
immediately surrounding the Galactic center because the broad \HI\
line emission contributes to the ATCA system temperature.  Here we
present only the part of the \HI\ data cube pertaining to the
Riegel-Crutcher cloud, the full SGPS GC dataset will be published in a
future paper.

\section{Results}
\label{sec:results} 
\begin{figure}
\centering
\includegraphics[width={\textwidth}]{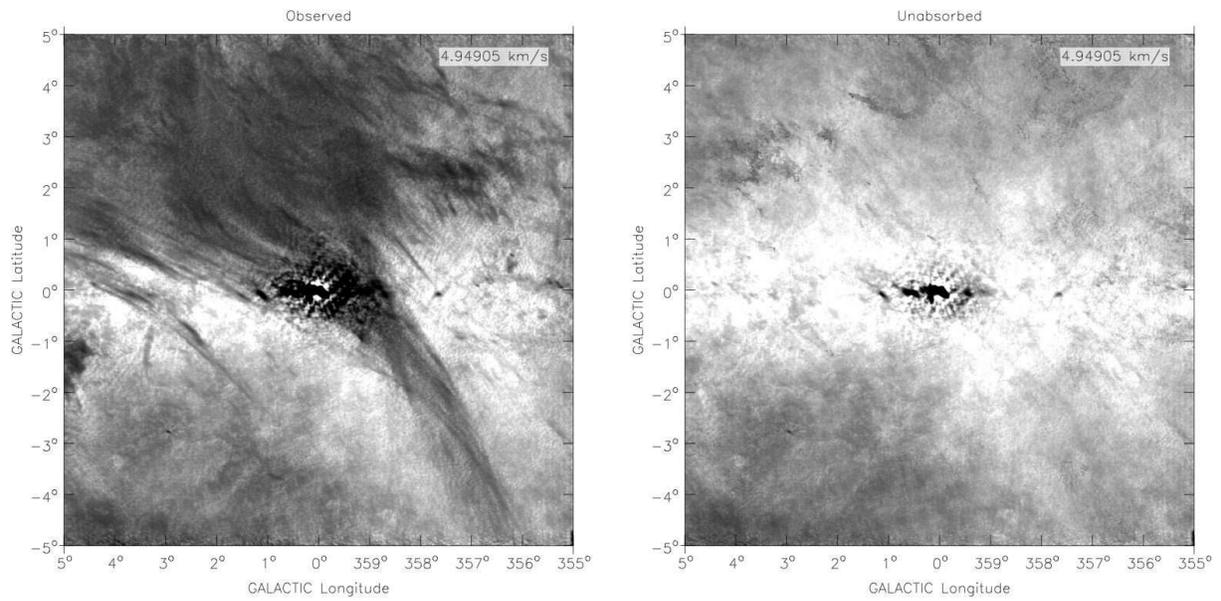}
\caption[]{SGPS \HI\ image at $v=4.95$ \kms\ towards the Galactic
  center.  The left panel shows the observed emission and the right
  panel shows the interpolated estimation of the unabsorbed emission.
  The grey scale is the same in both images, ranging from 15 K (black)
  to 140 K (white).  The angular resolution of these images is 100 arcseconds.
\label{fig:rc}}
\end{figure}

\begin{figure}
\centering
\includegraphics[angle=-90,width={\textwidth}]{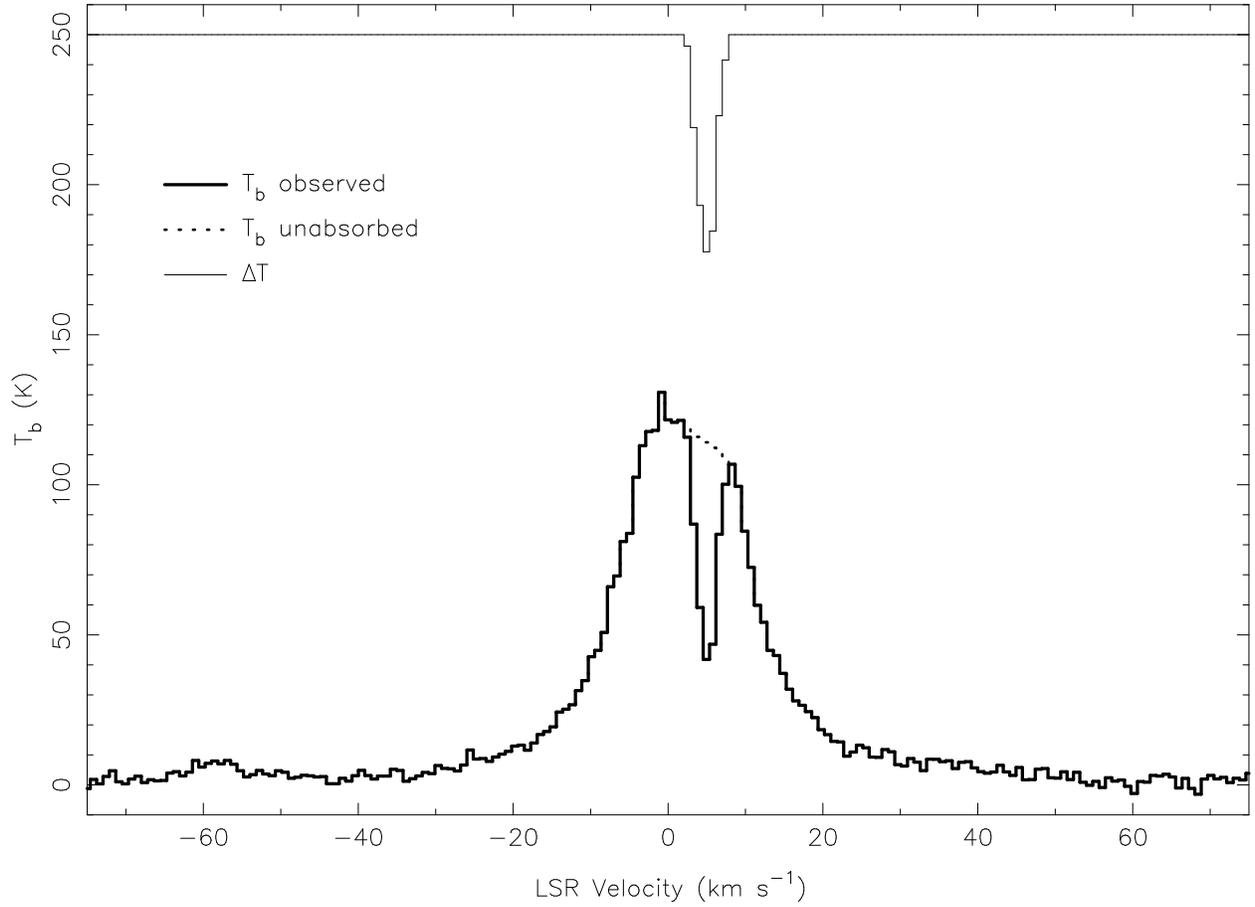}
\caption[]{\HI\ spectra towards $l=0\fdg79$, $b=1\fdg99$.  The heavy
  solid line shows the observed spectrum, $T_{\rm on}$, the dotted
  line is the interpolated spectrum, $T_{\rm off}$, and the thin solid
  line is the difference $\Delta T = T_{\rm on} - T_{\rm off}$.  The
  baseline of the difference spectrum has been offset by 250 K.  
  \label{fig:spec}}
\end{figure}

An \HI\ image at an LSR velocity of $4.95$ \kms\ from the final,
combined \HI\ data cube is shown in the left panel of
Figure~\ref{fig:rc}.  The Riegel-Crutcher (R-C) cloud is visible as
the large black swath that extends from upper left to the bottom right
of the image.  In order to estimate the properties of the cloud
itself, and to better examine its structure, we need to separate it
from the background \HI\ emission.  There are many ways of estimating
the unabsorbed \HI\ emission, $T_{\rm off}$, including using averages
of spectra observed adjacent to the cloud or interpolating across the
absorption with a linear, parabolic or high-order polynomial fit.
Each technique has its limitations; for very extended features like
the R-C cloud, off cloud spectra are too distant from the on-cloud
positions to accurately reflect the unabsorbed emission.  High order
polynomial interpolations across the absorption profile can give good
estimates for $T_{\rm off}$, but these are computationally expensive
for many spectra.  Depending on the line profile a simple linear fit
can also give a reasonable approximation to the off-cloud emission,
with minimal computational effort.  For simplicity we have chosen to
interpolate across the absorption with a linear fit.  An example of an
interpolated spectrum and the difference between the absorbed and
unabsorbed spectra, $\Delta T \equiv T_{\rm on} - T_{\rm off}$, is
shown in Figure~\ref{fig:spec}.  For each pixel in the GC cube with an
absorption depth of more than 8 K we searched for the edges of the
absorption feature, interpolated between those edges and replaced the
intermediate channels with the interpolated spectrum.  The result is a
cube of unabsorbed emission, $T_{\rm off}$.  The right panel of
Fig.~\ref{fig:rc} shows the unabsorbed emission for the same velocity
channel as the left panel.  Although the interpolation is not perfect,
it has removed the bulk of the R-C cloud.

\begin{figure}
\centering
\includegraphics[width=5.2in]{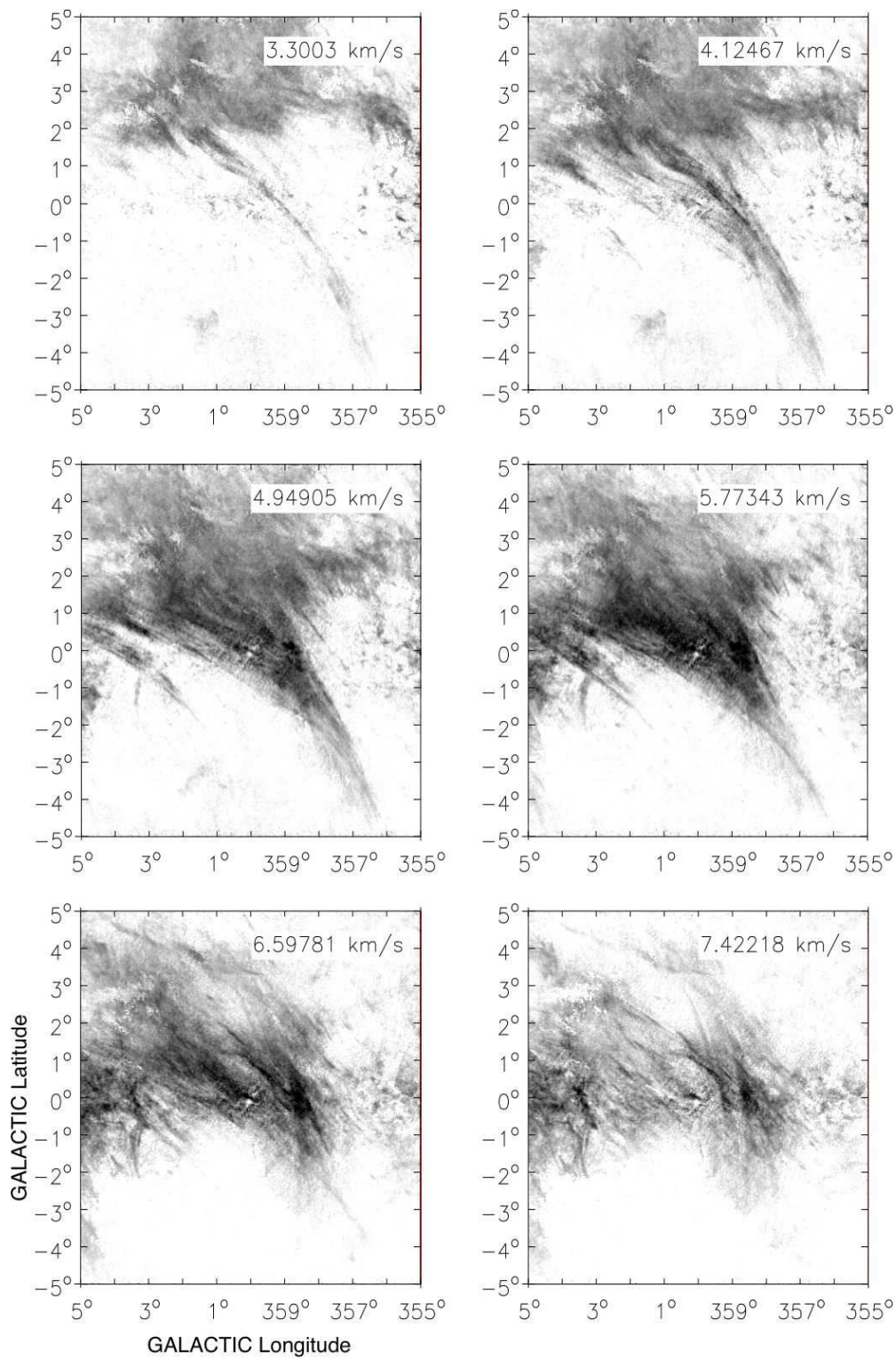}
\caption[]{Velocity channel images of $\Delta T$, the difference
  between the data and the interpolation, of the Riegel-Crutcher
  Cloud.  Every channel from $3.3$ \kms\ to $7.4$ \kms\ is shown.  The
  LSR velocity of the channel is shown in the upper right corner of
  the image.  The grey scale is linear between 0 (white) and -120 K
  (black).
  \label{fig:chans}}
\end{figure}

Velocity channel images of the difference, $\Delta T$, between the
observed and the unabsorbed cubes are shown in Fig.\ \ref{fig:chans}.
Here we show only velocity channels between 3.3 \kms\ and $7.4$ \kms,
which cover the majority of the R-C cloud at these longitudes.  The
images have a channel separation of $0.82$ \kms.  The bulk of the
cloud is observed in the top-left corner of the image, from there the
cloud extends to higher longitudes and latitudes, not covered by the
SGPS.  In this region the cloud appears fairly smooth in structure.
Extending from the bulk of the cloud towards the lower right of the
image is a wispy elongated structure.  Unlike the dense top of the
cloud, the tail appears to be resolved into a network of narrow
filaments, or strands, all roughly aligned.  We emphasize that the
edges of these filaments are real self-absorption features and are not
simply places where the background emission drops away.  The ensemble
of filaments has a distinct curvature to it and resembles a rope of
gas made up of individual strands.  The individual strands appear to
be mostly unresolved or in some cases marginally resolved.  They have
widths of between 2\arcmin\ and 5\arcmin, which, at a distance of
$125$ pc, correspond to physical widths of $0.07 - 0.2$ pc.  Some
filaments can be traced across the majority of the image; the longest
continuous filament is $7\fdg7$, or $16.9$ pc. Many of the filaments
have aspect ratios in the range 50-170:1.

\subsection{Properties of the Cloud}
\label{subsec:tempstruc} 
Deriving the temperature and density of an \HI\ cloud is non-trivial.
In theory these can be derived from solutions of the radiative
transfer equation.  In practice, though, the radiative transfer
equation cannot be solved for the complicated \HI\ profiles observed
in the Galaxy.  To derive the spin temperature, $T_s$, and optical
depth, $\tau$, of the R-C cloud we must make some simplifying
assumptions.  Here we use the four-component model outlined in
\citet{gibson00} and \citet{kavars03}.  This model assumes that all of
the continuum emission, $T_c$, originates behind the HISA cloud, but
allows for both foreground and background \HI\ with spin temperature
and optical depths of $T_{s,{\rm fg}}$, $\tau_{\rm fg}$ and $T_{s,{\rm
    bg}}$, $\tau_{\rm bg}$ respectively.  In this model, the radiative
transfer equation can be solved for the difference between the
observed brightness temperature on, $T_{\rm on}$, and off the cloud,
$T_{\rm off}$.  In the difference between on and off cloud the direct
contribution to the brightness temperature from the foreground cloud
is the same and cancels out.  The on-off difference is given by:
\begin{equation}
\Delta T \equiv T_{\rm on} - T_{\rm off} = \left[ T_s - T_{s,{\rm 
      bg}}\,\left(1-e^{-\tau_{\rm bg}}\right) - T_c\right] \,\,
      \left(1-e^{-\tau}\right) .
\label{eq:onoff}
\end{equation}
All variables, with the exception of $T_c$, are functions of
velocity, $v$.  We make a further assumption that $\tau_{\rm fg}$ and
$\tau_{\rm bg}$ are small.  As in \citet{gibson00} we use the variable,
$p$, to describe the fraction of the \HI\ emission originating
behind the self-absorption cloud.  These assumptions allow us to solve
equation (\ref{eq:onoff}) for the optical depth,
\begin{equation}
\tau = - \ln \left(1 - \frac{\Delta T}{T_s - T_c - pT_{\rm off}}\right) ,
\label{eq:tau}
\end{equation}
for a given value of $p$ in the range $0 \leq p \leq 1$.

The continuum temperature, $T_c$, is measured from our Parkes
continuum maps.  Although these are at a much lower resolution than
the ATCA data, the variation on small scales is only $10\%$ of the
mean continuum emission for a given latitude and not deemed important
for the rough temperature analysis presented here.  Included in equation
(\ref{eq:tau}) are the parameters $T_{\rm off}$ and $p$, which are not
directly observed and must therefore be estimated.  For $T_{\rm off}$
we use the interpolated cube described above.  Although the technique
of linear interpolation across the absorption feature can
underestimate the absorption in the line wings, we find that the
errors introduced are small when compared to $\Delta T$.  As our
analysis does not rely heavily on accurate linewidths, this does not
significantly affect our results.  To estimate $p$ we assume that
because the R-C cloud is at a distance of only 125 pc and lies just
beyond the edge of the Local Bubble \citep{crutcher84}, there is very
little foreground \HI\ emission and the majority of the \HI\ emission
will be behind the cloud.  We therefore assume that $p$ is constant
across the cloud and that it is equal to one, which will give a lower
limit for the optical depth.  We discuss the implications on $\tau$
for non-unity values of $p$ in \S \ref{subsec:caveat}.

It is clear from equation (\ref{eq:tau}) that the optical depth and
spin temperature of the cloud are degenerate.  Numerous authors have
explored methods of constraining either $T_s$ or $\tau$, including
some efforts applied specifically to the R-C.  For example,
\citet{montgomery95} estimated $T_s=35$ K by assuming that all of the
measured linewidth, $\Delta v$, for the narrowest absorption line can
be attributed to random kinetic motions, so that $T_s = T_k = 21.6 \,
(\Delta v)^2 \, {\rm K}$.  This equation must be used with care,
however, because it ignores all turbulent motions, which is not always
an appropriate assumption.  Spin temperature values estimated this way
should be considered only as upper limits.  This method also demands an
accurate measurement of the line-width, which as \citet{levinson80}
caution, can be difficult to attain for an \HI\ self-absorption
feature.

\begin{figure}
\centering
\includegraphics[angle=-90,width=5.5in]{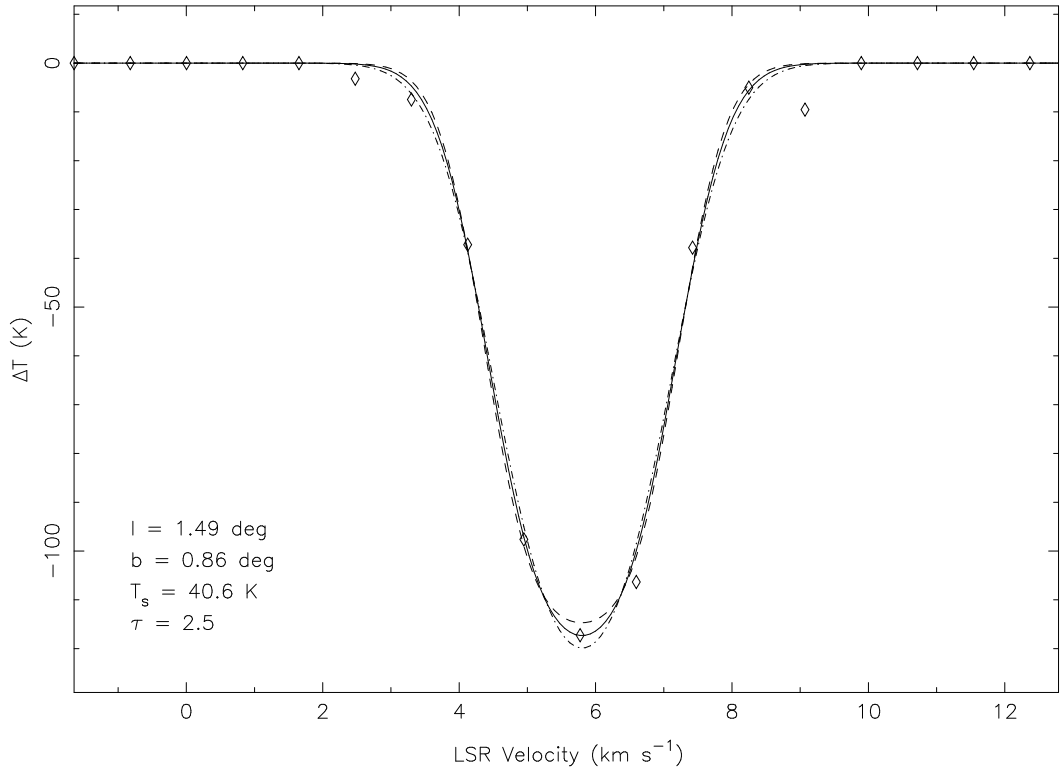}

\caption[]{\HI\ absorption spectrum at $(l,b) =  (1\fdg49,0\fdg86)$
  towards the R-C cloud.  The data points are shown with diamonds.  
  The fit to the marginally saturated line shape given by
  $T_s = 40$ K and $\tau = 2.5$ is plotted with the solid line.  For
  comparison, fits for $\tau = 3.5$, $T_s=51$ K and $\tau=1.5$,
  $T_s=14$ K are shown with dashed and dashed-dotted lines. Fits for
  values of $\tau < 1$ result in non-physical spin temperatures.  The
  $1-\sigma$ errors on the individual points are approximately the
  size of the plotted diamond.  

  \label{fig:taufit}}
\end{figure}

In some circumstances the line shape itself can be used to decouple $T_s$
and $\tau$.  Assuming a Gaussian optical depth of width $\Delta v$,
centered at $v_c$,
\begin{equation}
\tau(v) = \tau_0 \exp{[-2\ln 4(v-v_c)^2/\Delta v^2]},
\label{eq:tau2}
\end{equation}
the absorption profile defined by equation (\ref{eq:onoff}) will
flatten in the core as $\tau > 1$.  In this equation $\Delta v$ is
related to the line dispersion, $\sigma$, as $\Delta v =
\sqrt{8\ln2}\, \sigma$. To accurately trace the line shape requires
high signal-to-noise absorption spectra \citep{levinson80}, which are
not generally available. Fortunately, the depth of the absorption
towards the R-C cloud is so large that our observed absorption spectra
have S/N ratios in excess of 80, which makes it possible to fit
slightly flattened profiles.  There are several regions in the R-C
cloud where we observe slight saturation in the line profiles.  An
example is shown in Fig.\ \ref{fig:taufit}, where the line profile is
slightly non-Gaussian near the core, showing signs of saturation.  By
assuming a value for $p$, we can fit the saturated profiles to
uniquely measure $\tau$ and $T_s$.  Fitting this profile to equation
(\ref{eq:onoff}), where $\tau$ is given by equation (\ref{eq:tau2}),
we find $\tau = 2.5$ and $T_s = 40$ K.  The fit is shown in Fig.\
\ref{fig:taufit} as the solid line.  Fitted lines for constant values
of $\tau = 3.5$ and $\tau = 1.5$ are also overlaid for comparison.
Fitting all saturated profiles in the GC dataset we find that the spin
temperatures lie in the range 30 - 65 K.  Because of the errors
involved in this method for deriving spin temperatures these are only
estimates of the temperature. Comparing the range of spin temperatures
that we estimate with those estimated by \citet{montgomery95} and
\citet{crutcher84} we have chosen to adopt a spin temperature of $T_s
= 40$ K for the R-C cloud.  It is heartening to note that this is the
same value that \citet{heeschen55} found in his original work on this
feature.

The fit to the line profile also gives us an estimate for the observed
line width, $\sigma_{obs}$.  We assume that the measured linewidth is
related to the thermal and turbulent linewidths by $\sigma_{obs} =
\sqrt{\sigma_{turb}^2 + \sigma_{th}^2}$.  We find typical values
across the cloud of $\sigma_{obs}\sim 1.5$ \kms, which corresponds to a
line FWHM of $\Delta v = \sqrt{8\ln 2} \,\sigma_{obs}\sim 3.5$ \kms.  For
40 K gas the thermal linewidth is $\sigma_{th} \sim 0.6 $ \kms, which
allows us to approximate the turbulent linewidth as $\sigma_{turb}
\sim 1.4$ \kms.

\subsubsection{Density and Pressure}
\label{subsec:dens}
\begin{figure}
\centering
\includegraphics[width={\textwidth}]{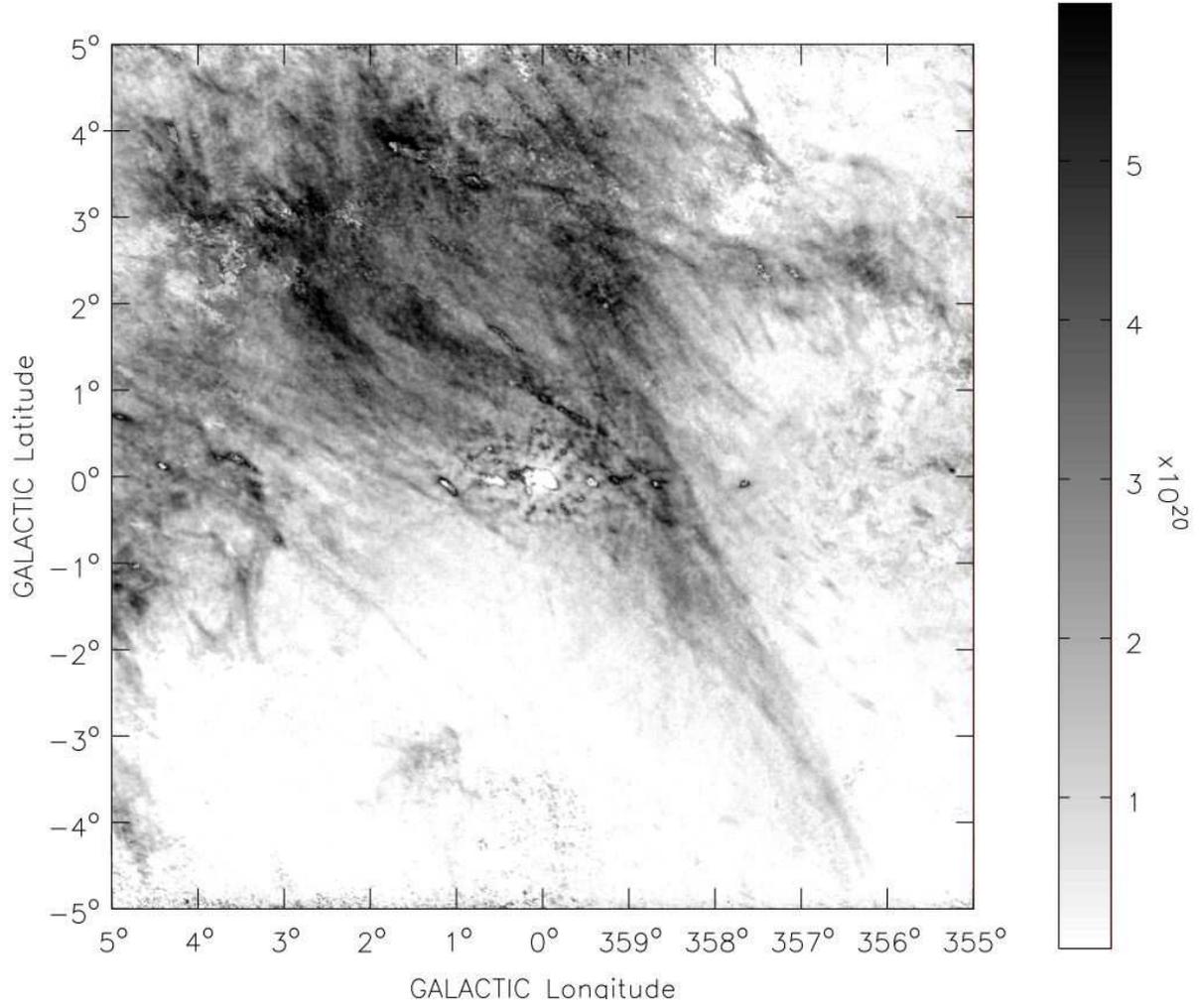}
\caption[]{\HI\ column density of the R-C cloud assuming a spin
  temperature of $T_s= 40$ K.  The grey scale is
  displayed in the wedge at the right in units of $10^{20}~{\rm cm^{-2}}$.
\label{fig:nhmap}}
\end{figure}

The column density of the \HI\ line is related to the spin temperature
and optical depth by:
\begin{equation}
N_{HI}=1.83\times10^{18} \, T_s \int \tau(v) dv \,\,\, {\rm cm^{-2}} ,
\label{eq:nh}
\end{equation}
which is valid for small $\tau$.  Using a constant value for $T_s$
across the cloud we can solve equation (\ref{eq:tau}) for the optical
depth as a function of position and velocity, which may be integrated
to produce a column density image for the R-C cloud.  We use a constant
value of $T_s=40$ K across the field.  An image of column density in
the R-C cloud is shown in Figure \ref{fig:nhmap}.  The mean column
density over the entire structure is $2.0 \pm 1.4 \times 10^{20}~{\rm
  cm^{-2}}$.  The densest regions are in the body of the cloud at
positive latitudes and longitudes.  There we measure column densities
of $\sim 4 \times 10^{20}~{\rm cm^{-2}}$, whereas the individual
filaments have lower column densities of $\sim 1 \times 10^{20}~{\rm
  cm^{-2}}$.

These estimates for the column density and spin temperature allow us
to make some order-of-magnitude estimates for the density and thermal
pressure of the cloud, assuming a cloud thickness, $\Delta s$.
\citet{crutcher84}, based on stellar absorption measurements, estimate
that the thickness of the R-C cloud is 1-5 pc.  However, our images
show that the individual filaments have plane-of-sky widths of less
than $\sim 0.1$ pc.  It is therefore possible that the strands are
cylindrical and that the thickness of the strands is similar to the
strand width, $\Delta s \sim 0.1$ pc.  We therefore use $\Delta s=0.1$
pc to estimate the \HI\ number density, $n_H$, and thermal pressure,
$P_{th}/k = N_{HI}T_s / \Delta s = n_H T_s$ in the individual
filaments.  The mean \HI\ density and pressure of the filaments are
$460~{\rm cm^{-3}}$ and $1.8 \times 10^4~{\rm K~cm^{-3}}$,
respectively.  The cylindrical geometry for the strands is more
intuitive than a collection of edge-on ribbons.  Nevertheless, the
data do not exclude an edge-on ribbon scenario.  If the filaments are
edge-on sheets of width $\Delta s \sim 1$ pc, as suggested from the
lower limit of \citet{crutcher84} then the mean \HI\ density and
pressure of the filaments are $46~{\rm cm^{-3}}$ and $1.8 \times
10^3~{\rm K~cm^{-3}}$, respectively.  For the upper left of the cloud we
suggest that the smoothness is indicative of many filaments integrated
along the line of sight.  We therefore assume that the width of the
base is $1-5$ pc, as estimated by \citet{crutcher84}.  Assuming
$\Delta s = 1 - 5$ pc, the \HI\ density in the base is $\sim 25 -
130~{\rm cm^{-3}}$, and the pressure is $(1 - 5)\times 10^3~{\rm
  K~cm^{-3}}$.

The thermal pressure of the filaments in the R-C cloud is almost an
order of magnitude larger than that expected for the CNM near the Sun,
which is believed to be only $P_{th}\sim 4000~{\rm K~cm^{-3}}$
\citep{wolfire03}.  While it is at first disconcerting to estimate
pressures that are so far out of thermal equilibrium, these pressures
are not exceptional.  \citet{jenkins01} found that as much as 25\% of
the neutral carbon bearing ISM has thermal pressures in the range
$P/k= 10 ^{3.5 - 4.0}~{\rm K~cm^{-3}}$ and with a few percent having
pressures larger than $P/k= 10 ^{4.0}~{\rm K~cm^{-3}}$.  In most areas
of the ISM the thermal pressure is a small fraction of the total
pressure.  We can estimate the total pressure in the R-C cloud as $P =
\sigma^2 \rho$ including both the thermal and turbulent motions, where
$\rho$ is the mass density assuming 10\% Helium, $\rho = 1.4m_Hn_H$.
For $\sigma \approx 1.5~{\rm km~s^{-1}}$ and an average number density
of $n_H \sim 460~{\rm cm^{-4}}$, the total pressure in the filaments
is $P/k \sim 2\times 10^5~{\rm K~ cm^{-3}}$. Again the total pressure
is about a factor of ten larger than total pressure (thermal plus all
non-thermal components) for the general ISM in the midplane
\citep{boulares90}.

\subsubsection{Density and Pressure Caveats}
\label{subsec:caveat}
It is worth taking note of some of the caveats to the density and
pressure estimates given above.  These estimates rely on assumptions
made about $T_s$, $p$, and $\Delta s$.  Our column density image
assumes a constant spin temperature across the cloud.  This is almost
certainly not true, but in the absence of saturation in the \HI\
profiles it is not possible to solve for $T_s$ and $\tau$ for all
positions.  Letting $T_s$ vary randomly around the image about a mean
40 K with a variance of 10 K we found that the densities measured do not
vary significantly from those determined here.   

The exact value of $T_s =40$ K assumed for the spin temperature has an
effect on the densities and pressures derived.  The cosmic microwave
background sets an absolute minimum value for $T_s\geq 2.73$ K, which
results in column densities about an order of magnitude smaller than
those derived for $T_s=40$ K.  Such a low spin temperature is highly
unlikely in almost all Galactic environments where the UV radiation field
and cosmic ray heating raise the temperature well above the CMB
background.  If we assume that $T_s =30$ K, which is at the lower end
of values estimated by previous authors, then the derived column
densities are a factor of $\sim 0.6$ smaller than for $T_s = 40$ K.
Values of $T_s>40$ K result in undefined values of $\tau$ towards
regions where $\left|\Delta T\right|$ is large.

The assumed value of $p$ can significantly effect the derived optical
depth and column density.  By assuming $p=1$ we have minimized the
optical depth for any given value of $T_s$ and therefore minimized the
densities and pressures calculated.  In addition, values of $p\approx1$
are most likely given the close proximity of the cloud.  If we
decrease $p$ to an unlikely value of $0.8$ it has two main effects:
first, the optical depth becomes undefined over large areas of the map
and second, the derived column densities are typically a factor of two
larger than for $p=1$.

Finally, the assumed cloud thickness has a large impact on the derived
densities and pressures.  We have assumed a thickness for the
filaments that is an order of magnitude smaller than previous
estimates by \citet{crutcher84}. However, \citet{crutcher84} used
stellar absorption measurements at distances bracketing the cloud to
estimate its thickness.  Those measurements did not have subparsec
precision and they would not have been able to exclude a thickness of
less than 1 pc.  At the other extreme, if we adopt the
\citet{crutcher84} upper limit of $\Delta s = 5$ pc for the filaments,
the derived pressures and densities will be a factor of fifty smaller
than those derived for $\Delta s \sim 0.1$ pc.  This would present a
very unusual ribbon-like geometry for the R-C filaments, which we do
not believe is likely.

\section{The  Magnetic Field Structure }
\label{sec:bfield} 
\begin{figure}
\centering
\includegraphics[width={\textwidth}]{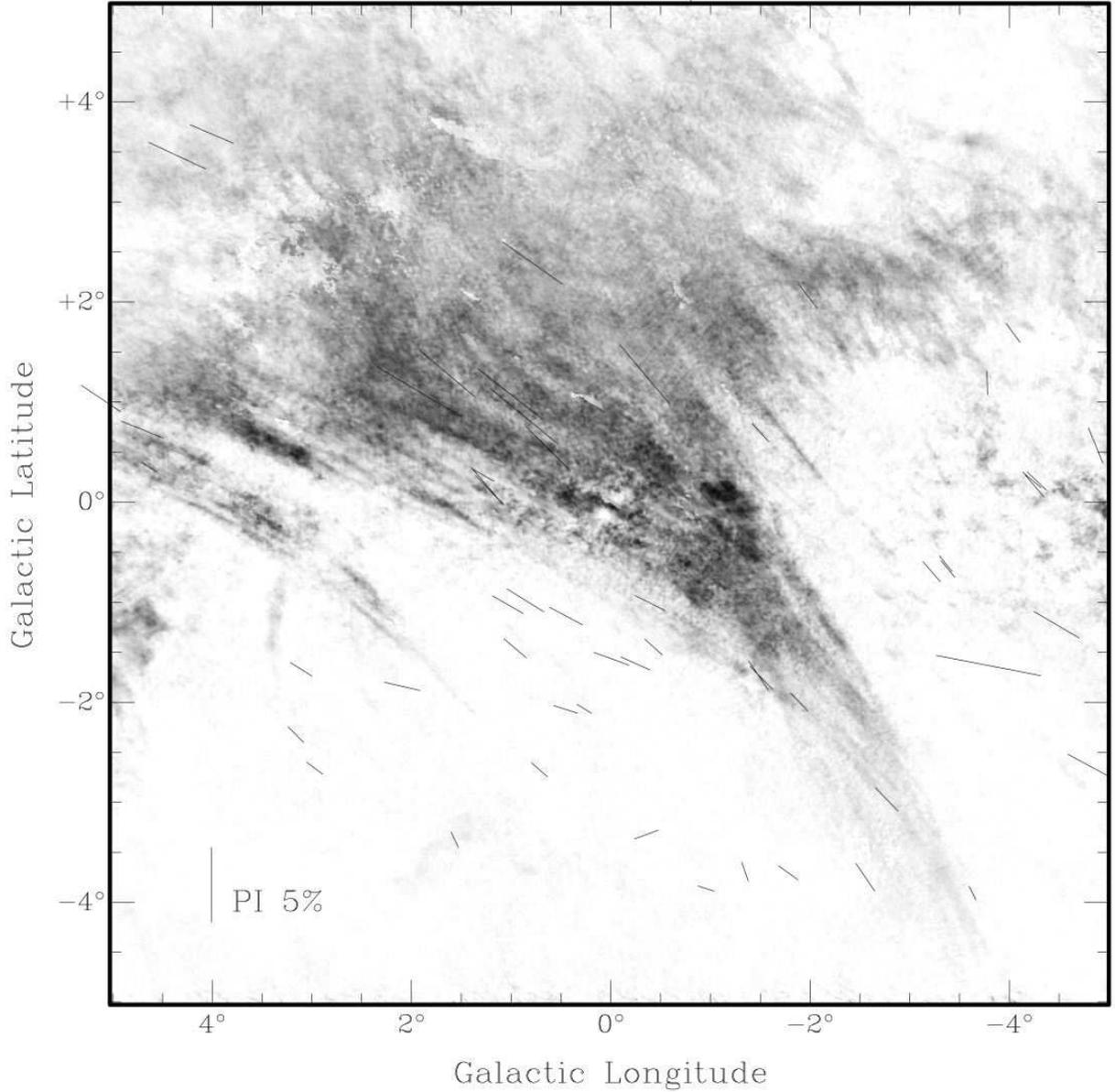}
\caption[]{\HI\ image of the R-C cloud at $v=4.95$ \kms\ overlaid with
  vectors of stellar polarization from \citet{heiles00}.
  The measured polarization vectors are aligned with the magnetic
  field direction.  The  length of the vectors is proportional to the measured fractional polarized
  intensity, with the scale given by the 5\% fractional polarized intensity vector shown by the scale of the vector in the lower
  left corner.    
  \label{fig:polmap}}
\end{figure}

The filaments observed here, though curved, have virtually no wiggles.
This taughtness suggests that the gas structure may be dominated by
some process other than turbulence, such as a magnetic field. To
explore this suggestion we have searched for the orientation of the
magnetic field associated with the R-C cloud. We used the
\citet{heiles00} compilation of optical measurements of stellar
polarization to determine the magnetic field direction. From
\citet{heiles00} we extracted all stars at distances of less than 2
kpc in the region $-5\arcdeg \leq l \leq +5\arcdeg$, $-5\arcdeg \leq b
\leq +5\arcdeg$.  To ensure reliable polarization angle measurements
we further constrained the selection to stars with polarized intensity
greater than 1\% of the total intensity; 56 stars met the criteria.
In Figure~\ref{fig:polmap} we have overlaid polarization vectors on
the \HI\ $\Delta T$ image of the R-C cloud at $v=4.95$ \kms.  Vectors
are shown at the position of each polarized star, oriented with the
polarization angle,\footnote{The polarization angles are measured such
  that $\theta_p=0\arcdeg$ when the vector points towards up and
  increases in a clockwise direction} $\theta_p$.  The length of the
vectors is proportional to the measured fractional polarized intensity
of the star, with a vector of 5\% fractional polarization shown in the
key at the bottom-left.  In the regions covered by the cloud there is
a remarkable agreement between the orientation of the polarization
vectors and the direction of the cloud's filaments.  By contrast, in
the area $l\geq -2\arcdeg$ and $b\leq -3\arcdeg$ there is very little
absorption associated with the cloud and these regions show disordered
polarization vectors.  Referring only to the area covered by the cloud
($l>-3\arcdeg$, $b>-3\arcdeg$) the mean polarization angle is
$\left<\theta_p\right> = 53\arcdeg \pm 11\arcdeg$.  The polarization
angle for polarization arising from polarization of background
starlight is parallel to the magnetic field direction such that the
angle of the magnetic field, $\theta_B$, also equals $53\pm11\arcdeg$.

Stellar polarization measurements probe the magnetic field integrated
along the line of sight, which can lead to a superposition of
structures in the observed vectors.  Because we initially included all
stars out to 2 kpc distance it is also possible that the magnetic
field orientation observed is not associated with the R-C cloud, but
located behind it.  To test this we examined only stars out to a
distance of 200 pc and found that, although the sample size is
smaller, the mean polarization angle agreed with
$\left<\theta_p\right> = 53\arcdeg$ to within the standard deviation
of the full sample.  This suggests that the dominant magnetic
structure along this line of sight lies within 200 pc and the very
good alignment with the R-C \HI\ strands suggests that they are
related.

\section{Discussion}
\label{sec:discussion} 
\subsection{Magnetic Field Strength}
\label{subsec:bstrength}
Measurements of stellar polarization do not directly yield a
measurement of the strength of the magnetic field.  However,
\citet{chandrasekhar53} suggested that the variance in an ensemble of
polarization measurements can be used to estimate the strength of the
field.  This method is often applied to measurements of starlight
polarization due to dust \citep[e.g.][]{andersson06} and has been
tested against theoretical MHD models by \citet{heitsch01} and
\citet{ostriker01}.  The technique assumes that turbulence in the
magnetized medium will randomize the magnetic field and that the
stronger the regular field is, the less it is disturbed by turbulence.
Using the Chandrasekhar-Fermi (C-F) method, as modified by
\citet{heitsch01}, the strength of the magnetic field in the plane of
the sky can be estimated as:
\begin{equation}
\left< B \right>^2 = \xi \,4 \pi \rho \frac{\sigma_v^2}{\sigma (\tan
  \delta_p)^2},
\label{eq:cf}
\end{equation}
where $\sigma_v$, is the turbulent linewidth for the R-C filaments,
$\rho$ is the density of the medium, $\delta_p \equiv \theta_p -
\left<\theta_p\right>$, and $\xi$ is a correction factor
\citep[e.g.][]{heitsch01}. For small values of $\delta_p$,
$\sigma(\tan \delta_p) \approx \sigma(\delta_p)$.  The derived
magnetic field depends on the assumed value for $\xi$.
\citet{heitsch01} found that for simulated molecular clouds the value
of $\xi$ required to reconcile the C-F estimated field strength and
the input field was in the range $0.2 - 1$, with an average value of
0.5 for the ensemble of simulated clouds.  \citet{ostriker01} also
found an average correction factor of 0.5 for simulated molecular
clouds, although with a smaller dispersion.

To estimate the magnetic field strength in the R-C cloud we will adopt
$\xi =0.5$.  The dispersion in the measured stellar polarization
angles towards the R-C cloud is $\sigma(\delta_p) \approx 11\arcdeg$
and, as before, $\rho = 1.4 m_H n_H = 1.1 \times 10^{-21}{\rm
  g~cm^{-3}}$ and $\sigma_v = \sigma_{turb} = 1.4$ \kms. We therefore
estimate that $\left <B\right> \sim 60~{\rm \mu G}$. Because of the
uncertainty in the correction factor, $\xi$, for a given cloud, the
field strength estimated from the C-F method must be regarded as a
rough estimate, probably only good to within a factor of two.

As an alternative to the C-F method, we can estimate the magnetic
field strength assuming equipartition of the magnetic and kinetic energy
densities in the cloud.  The R-C cloud filaments appear to be very
straight compared with other structures observed in the ISM.  Magnetic
tension can provide a mechanism for holding filaments straight against
turbulent effects.  For the magnetic field to dominate the structure
it must dominate over internal turbulence and gravitational effects.
We can estimate the relative contributions of the kinetic,
gravitational, and magnetic energy densities, ${\mathcal K, G, M}$,
respectively.  The kinetic energy density for an \HI\ cloud is given
by
\begin{equation}
{\mathcal K} = 3/2 \, (1.4 \, m_H) \, n_H \, \sigma^2 = 2.1 \times
10^{-33} \, \frac{N_H \Delta v^2}{\Delta s}~ {\rm ergs~cm^{-3}}, 
\end{equation}
where $\Delta v$ is in ${\rm km~s^{-1}}$ and $\Delta s$ is in
pc. Considering just the filaments of the R-C cloud, where $N_H \sim 1
\times 10^{20}~{\rm cm^{-2}}$, $\Delta v \sim 3.5$ \kms, and $\Delta s
\approx 0.07$ pc if the filaments are cylinders, then the kinetic
energy density is ${\mathcal K} = 3.7 \times 10^{-11}~{\rm
  ergs~cm^{-3}}$.  If the filaments are edge-on ribbons instead of
cylinders then the kinetic energy density will be smaller.

If we approximate the filaments as cylinders of $r=\Delta s/2$ and
length $L$, then the total gravitational energy is given by $W = - G
M^2 / L$ \citep{fiege00} and the gravitational energy density is:
\begin{equation}
{\mathcal G} = G \pi (1.4m_H\, n_H\, r)^2,
\end{equation}
For the R-C filaments, the gravitational energy density is ${\mathcal
  G} \sim 3 \times 10^{-15}~{\rm ergs~cm^{-3}}$.  The gravitational
energy density is very small compared to the kinetic energy density;
these filaments are clearly not self-gravitating.

Finally, the magnetic energy density is ${\mathcal M} =
|B_{tot}|^2/8\pi$, where $B_{tot}\geq B_{meas}$.  Given negligible
gravitational energy density, for the magnetic field to dominate the
structure observed in the R-C cloud, ${\mathcal M} > {\mathcal K}$.
For ${\mathcal K} \sim 3.7 \times 10^{-11}~{\rm ergs~cm^{-3}}$ the
magnetic field strength must be $B_{tot}> 30~{\rm \mu G}$.  The
agreement between the magnetic field strength derived assuming
magnetic dominance and with the C-F method is surprisingly good.  One
can use this result and Eq.~\ref{eq:cf} to solve for $\xi$ for the R-C
cloud. A field strength of $> 30~{\rm \mu G}$ implies that the
correction factor, $\xi$, for this cloud should be $>0.12$.

To better understand the magnetic field of the R-C cloud we would also
require measurements of the line-of-sight magnetic field strength.
Most measurements of magnetic fields in the CNM have been made using
Zeeman splitting \citep[e.g][]{heiles05}, which probes the
line-of-sight component of the field. \citet{heiles05} find a median
CNM magnetic field strength of $6.0\pm1.8~{\rm \mu G}$ over a wide
range of Galactic environments.  This is considerably lower than the
values estimated here for the R-C cloud.  It is important to note,
however, that the density of the R-C cloud is larger than typical CNM
values.  In fact, the magnetic field strength estimated is more
consistent with values obtained for molecular clouds with densities
similar to that of the R-C cloud \citep{crutcher99}.  However, for
these measurements it is usually assumed that gravitational
contraction compresses the magnetic field lines leading to an increase
in density and magnetic field strength.  There is no evidence for such
action in the R-C cloud and it is therefore curious that it should
exhibit such a high magnetic field strength.

\subsection{Comments on the Structure of the CNM}
\label{subsec:cnm}
The structure of the cold neutral medium (CNM) is very difficult to
study. For many years we have harbored the notion that the CNM was
distributed in isotropic clouds as described in the \citet{mckee77}
paradigm.  However, there is a fair amount of evidence against this
model.  For example, high resolution \HI\ absorption measurements
towards pulsars and radio galaxies find very small-scale variations in
the \HI\ that seem to require unreasonably high thermal pressures if
we assume they are spherical.  \citet{heiles97} invoked geometry to
explain this ``tiny scale atomic structure'' (TSAS), suggesting
instead that these measurements probe cold gas in thin sheets viewed
edge-on or filaments viewed end-on.  Although the sheet or filament
description is appealing because of its resolution of the overpressure
problem, it has not been readily adopted, owing largely to the lack of
observational evidence for counterparts aligned perpendicular to the
line of sight.  These sheets or filaments are predicted to have very
low column densities, rendering them unobservable \citep{heiles97}.
There is evidence that on larger scales the CNM is indeed in
large-scale filaments of sheets.  In the few examples where we can
image large areas of the CNM through \HI\ self-absorption, such as in
the CGPS and SGPS \citep{gibson05,kavars05}, the CNM seems extended,
with complicated structure that is far from isotropic.  \HI\
absorption measurements towards closely separated continuum sources
support this view with similar absorption components that suggest
extended CNM clouds with aspect ratios $\sim 10 - 200$
\citep{heiles03b}.  With our images we find that the cold medium of
the R-C does not resemble the spherical clouds of \citet{mckee77} but
instead is structurally much closer to the \citet{heiles97}
description of filaments. Our filaments differ from those proposed by
\citet{heiles97} in that they are viewed side-on, rather than end-on.
Although the size scales represented by the R-C cloud are more than
two orders of magnitude larger than those observed in TSAS, the R-C
cloud presents evidence that the CNM on small scales is structured in
filaments with extreme aspect ratios.

\subsection{Comparison with Molecular Clouds}
\label{subsec:molecules}
As much as 60\% of the \HI\ self-absorption features observed in the
inner Galactic plane have some association with molecular gas as
detected by $^{12}$CO emission \citep{kavars05}.  Comparing the \HI\
data with data from the low resolution ($30\arcmin$) $^{12}$CO survey
\citep{dame87}, there is some indication for associated CO emission at
high latitudes around $v\approx 3.5$ \kms.  Without CO data of
comparable resolution to the \HI\ it is difficult to make a clear
association between the \HI\ self-absorption filaments and CO
emission.  Although CO emission in the dense base of the R-C cloud
seems reasonable, CO emission in the filaments would be surprising.
The filaments' column densities of $\sim 1 \times 10^{20}~{\rm
  cm^{-2}}$ should account for only $\sim 0.05$ mag of visual
extinction, which is below the extinction limit generally required to
shield CO from photo-dissociation \citep{vandishoeck88}.

Despite the lack of clear molecular emission in the filaments it may
be informative to compare the filamentary structure of the R-C cloud
with that seen in molecular clouds.  \citet{kavars05} suggest that gas
probed by \HI\ self-absorption may fill the evolutionary gap between
diffuse atomic clouds and molecular clouds.  The density of the R-C
cloud filaments, $n \sim 460~{\rm cm^{-3}}$ is much higher than most
\HI\ self-absorption features \citep{kavars05} and indeed most of the
CNM \citep{heiles03b}, it is instead closer to the low end of
molecular cloud densities \citep{crutcher99}.  Morphologically, the
R-C cloud is reminiscent of the structure observed in some molecular
clouds.  Molecular clouds have long been observed to have elongated
filamentary structure similar to that observed in the R-C cloud.  Good
examples of filamentary molecular clouds are the Taurus, $\rho$
Ophiuchus and Orion molecular clouds, which show long molecular
filaments with dense clumps along their lengths
\citep[e.g.][]{mizuno95,falgarone01}.  Although it is not clear how
filamentary structure develops in molecular clouds, it is expected
that magnetic fields play an important role in establishing and
maintaining their structure.

Far-infrared polarization observations of filamentary molecular clouds
show that very often the polarization vectors are oriented either
along or perpendicular to the filaments \citep{matthews00}.  This
bimodal distribution has often been attributed to helical magnetic
fields that wrap around the filaments \citep{carlqvist97,fiege00}.
However, reliably inferring the three dimensional magnetic field
structure from plane-of-sky dust polarization measurements is
difficult.  Some of the most promising evidence for helical magnetic
field structure has come from Zeeman measurements of Orion A
\citep{robishaw05}, which show the field reversing direction on
opposite sides of the filament.

A fundamental difference between filamentary molecular clouds and the
R-C cold \HI\ filaments is that molecular clouds are in general
self-gravitating.  Depending on the field geometry the magnetic field
in molecular clouds can provide crucial support against gravitational
collapse.  For example, for fields aligned along the direction of
elongation (poloidal fields) the magnetic field provides an outwards
pressure to counter the gravitational contraction.  For helical fields
the situation is more complex with the net pressure effect at the
surface of the cloud depending on the relative strengths of the
poloidal and azimuthal components.  For helical fields where the
azimuthal component dominates the magnetic pressure acts inwards.  

For the R-C cloud the true configuration of the magnetic field is
crucial to defining the cloud life-time.  It is also particularly
difficult to distinguish the field configuration within the cloud from
the configuration of the ambient field from the stellar polarization
alone.  If the field within the filaments is poloidal or helical with
a dominant poloidal component, as the observed polarization vectors
suggest, then the magnetic field has the effect of increasing the
total pressure.  The over-pressurized filaments should be short-lived
with radial crossing times on the order of $4\times 10^4$ yr.  If, on
the other hand, the field within the filaments is actually helical and
has places where the azimuthal field dominates, then the magnetic
pressure may stabilize against the thermal pressure and could even
lead to collapse and fragmentation of the filaments.  Unfortunately it
is impossible to infer the true internal magnetic field configuration
from the starlight polarization data of these multiple superimposed
filaments. Zeeman measurements of the line-of-sight magnetic field
component are needed to understand the dynamical future of the R-C
cloud.
\section{Conclusions}
\label{sec:conclusions}
We have presented new high resolution \HI\ self-absorption images of
part of the Riegel-Crutcher (R-C) cold cloud at a distance of 125 pc
in the direction of the Galactic center.  These images, which are part
of the SGPS Galactic Center survey, cover the area $-5\arcdeg \leq l
\leq 5\arcdeg$, $-5\arcdeg \leq b \leq 5\arcdeg$ with an angular
resolution of 100 arcseconds.  The data reveal extraordinary
filamentary structure within the R-C cloud.  The cloud seems to be
composed of a network of aligned narrow filaments that merge into a
smooth mass at the upper left of the cloud.  The filaments are very
elongated, with the most extreme example only $0.07$ pc wide and over
16 pc long.  Using saturated absorption profiles in the cloud we
estimate a spin temperature for the cloud of $T_s \sim 40$ K, which is
similar to values previously estimated by \citet{crutcher74} and
\citet{montgomery95}.  Adopting a spin temperature of 40 K for the
entire cloud we created images of the optical depth and column density
of the cloud, finding an average column density of $\sim 2 \times
10^{20}~{\rm cm^{-2}}$.  If the individual filaments of the cloud are
approximated as cylinders, then their average thermal pressure is
$\sim 2 \times 10^{4}~{\rm K~cm^{-3}}$.

From measurements of the polarization of starlight from
\citet{heiles00} we have shown that the filaments in the R-C cloud are
aligned with the magnetic field of the cloud.  Using the
Chandrasekhar-Fermi method we have roughly estimated the magnetic
field strength to be $\sim 60~{\rm \mu G}$.  By comparing ratios of
gravitational, kinetic and magnetic energy densities we show that the
filaments are not self-gravitating.  We suggest that the excellent
alignment between the filaments and the magnetic field, as well as the
straightness of the filaments, is evidence that the magnetic field
dominates over thermal and turbulent motions within the cloud. For the
magnetic energy density to exceed the kinetic energy density the
magnetic field must be $>30~{\rm \mu G}$.  Given the uncertainties
associated with the Chandrasekhar-Fermi method the two field
strengths are fully consistent.  

Directly imaging the structure of the CNM is always difficult.  The
R-C cloud, because it is in front of the bright, high column density
\HI\ background towards the Galactic center, provides us with a unique
opportunity to directly image the CNM.  In this case the CNM appears
highly filamentary and most closely resembles the high aspect ratio
filaments suggested by \citet{heiles97} to explain Tiny Scale Atomic
Structure.  There is no reason to believe that the structure of the
R-C cloud is unique, it may in fact be typical for the CNM.  We point
out that there are morphological similarities between the R-C cloud
and filamentary molecular clouds, which are also characterized by
strong magnetic fields.  If gas traced by \HI\ self-absorption is
either a precursor to molecular gas or formed from dissociation of
molecular clouds the morphological similarity is not surprising.

\acknowledgements The Parkes Radio Telescope and the Australia
Telescope Compact Array are part of the Australia Telescope which is
funded by the Commonwealth of Australia for operation as a National
Facility managed by CSIRO.  MH acknowledges support from the National
Radio Astronomy Observatory (NRAO), which is operated by Associated
Universities Inc., under cooperative agreement with the National
Science Foundation.  The authors thank an anonymous referee for
insightful and interesting questions and suggestions.


\clearpage
\begin{deluxetable}{ll}
\tablewidth{0pt}
\tabletypesize{\footnotesize}
\tablecaption{SGPS Galactic Center ATCA Observations
\label{tab:obs}}
\tablehead{
\colhead{Dates} & \colhead{Array Configuration} }
\startdata
2002 Dec 21 - 31 & EW367 \\
2003 Mar 24 - 31 & EW352 \\ 
2003 May 11 - 13 & EW352 \\
2003 Jun 27 - 29 & 750C \\
2003 Aug 11 - 15 & EW367 \\
2003 Sep 11 - 19 & 750C \\
2004 Feb 19 - Mar 01 & 750A \\
2004 Jun 02 - 14 & 750D \\
2004 Nov 13 & 750C\tablenotemark{a} \\
\enddata
\tablenotetext{a}{Make-up session.}
\end{deluxetable}

\clearpage

\end{document}